\documentclass[aps,twocolumn,amsmath,amssymb,floatfix]{revtex4}

\usepackage{graphicx}% Include figure files
\usepackage{dcolumn} % Align table columns on decimal point
\usepackage{bm}      % Bold math
\usepackage{color}

\def\ADD#1{{\textcolor{blue}{#1}}}         % addition
\begin{document}

\title{Large scale behavior and statistical equilibria in rotating flows}
\author{P.D. Mininni$^{1,2}$, P. Dmitruk$^1$, W.H. Matthaeus$^3$ and A. Pouquet$^2$}
\affiliation{$^1$Departamento de F\'\i sica, Facultad de Ciencias Exactas y
         Naturales, Universidad de Buenos Aires and IFIBA, CONICET, Ciudad 
         Universitaria, 1428 Buenos Aires, Argentina. \\
             $^2$Computational and Information Systems Laboratory, NCAR, 
         P.O. Box 3000, Boulder, Colorado 80307-3000, USA. \\
             $^3$ Bartol Research Institute and Department of Physics and 
         Astronomy, University of Delaware, Newark DE 19716, USA.}
\date{\today}

\begin{abstract}
We examine long-time properties of the ideal dynamics of three--dimensional flows, in the presence or not of an imposed solid-body rotation and with or without helicity (velocity-vorticity correlation). In all cases the results agree with the isotropic predictions stemming from statistical mechanics. No accumulation of excitation occurs in the large scales, even though in the dissipative rotating case anisotropy and accumulation, in the form of an inverse cascade of energy, are known to occur. We attribute this latter discrepancy to the linearity of the term responsible for the emergence of inertial waves. At intermediate times, inertial energy spectra emerge that differ somewhat from classical wave-turbulence expectations, and with a trace of large-scale excitation that goes away for long times. These results are discussed in the context of partial two-dimensionalization of the flow undergoing strong rotation as advocated by several authors.
\end{abstract}
\pacs{47.32.Ef, 05.20.Jj, 47.27.eb, 47.11.Kb}
\maketitle

\section{Introduction}

Geophysical and astrophysical flows are turbulent. The dimensionless parameter that measures such a state, the Reynolds number, is huge for the atmosphere and the ocean, as well as for the sun and stars, the interplanetary medium, and the interstellar medium. Nonlinear effects, e.g., through advection, have thus ample time to act; they develop strong localized intermittent structures such as vortex filaments (or current sheets when dealing with magnetic fields), and a myriad of modes are excited. One thus resorts to a statistical description of such flows and examine the behavior of probability density functions (PDFs) as well as scaling exponents of structure functions. At second order, one obtains a power-law for the distribution of energy among the Fourier modes, in a direct cascade of energy to small scales where the energy is finally dissipated \cite{Kolmogorov41}. Another striking feature of turbulent flows is that, in some cases (including in three space dimensions in the presence of rotation), large-scale modes are excited as well \cite{Kraichnan67,Batchelor69,Cambon97}.

A deep understanding of turbulence is still lacking, in part because of the large number of interacting modes due to the convolution in Fourier space stemming from the nonlinearities, quadratic in the incompressible case. Dimensional analysis and self-similarity in the isotropic case leads to the Kolmogorov energy spectrum of distribution of energy among Fourier modes $k$, which reads $E(k)\sim \epsilon^{2/3}k^{-5/3}$, with $\epsilon$ the energy injection rate. This law is rather well verified in the atmosphere \cite{Nastrom84} and in the solar wind \cite{Bavassano82,Matthaeus86}, in laboratory \cite{Frisch95,Antonia97} as well as when examining numerical experiments \cite{Kaneda03}. When considering the helicity spectrum $H(k)$, where the helicity is the correlation between the velocity and its curl, the vorticity, one finds $H(k)\sim k^{-5/3}$ (see for example \cite{Chen03,Chen03b,Mininni06}). However, it was recently shown numerically and phenomenologically that, in the presence of rotation, other spectra emerge. This is the consequence of the effect of the inertial waves together with the fact that the helicity dominates the direct cascade to the small scales, the energy mostly undergoing an inverse cascade to the large scales \cite{Mininni09a, Mininni10a} (see also  \cite{Brissaud73}). 

Finding an underlying low-dimensional attractor to the complex behavior of turbulence is one way that such flows could be understood in simpler terms, as e.g., coherent structures with super-imposed noise. Another possibility is to resort to statistical mechanics and the dynamics of an ideal system with a finite number of modes. Equilibrium statistical mechanics has proven valuable in that it could predict in many cases the direction of the transfers in wavenumber space, either to the small scales (direct) or to the large scales (inverse) \cite{tdlee,Fox73,Novikov74,Kraichnan75,rhk_montgo}.

\begin{table}
\caption{
Parameters of the runs: band of wavenumbers $k_0$ excited in the initial conditions, linear spatial resolution $N$, rotation rate $\Omega$, initial relative helicity $\rho$, Rossby number Ro, and micro Rossby number R$_\omega$ computed at $t=0$ (see text for definitions). At late times, R$_\omega$ is of order unity in all runs as a result of the thermalization.}
\begin{ruledtabular}
\begin{tabular}{ccccccccccc}
Run & $k_0$  & $N$ &$\Omega$&$\rho$& Ro     & R$_\omega$\\
\hline
R1  & 2-4    & 128 & 0      & 0    &$\infty$& $\infty$  \\
R2  & 2-4    & 128 & 16     & 0    & $0.02$ & $0.08$    \\
R3  & 2-4    & 128 & 16     & 1    & $0.02$ & $0.08$    \\
R4  & 2-4    & 128 & 32     & 0    & $0.01$ & $0.04$    \\
R5  & 2-4    & 128 & 32     & 1    & $0.01$ & $0.04$    \\
R6  & 2-4    & 128 & 64     & 0    & $0.005$& $0.02$    \\
R7  & 2-4    & 128 & 64     & 1    & $0.005$& $0.02$    \\
R8  & 2-4    &  64 & 16     & 0    & $0.02$ & $0.08$    \\
R9  & 2-4    &  64 & 16     & 1    & $0.02$ & $0.08$    \\
R10 & 2-4    & 256 & 8      & 0    & $0.04$ & $0.16$    \\
R11 & 2-4    & 256 & 16     & 0    & $0.02$ & $0.08$    \\
R12 & 2-4    & 256 & 16     & 1    & $0.02$ & $0.08$    \\
R13 & 2-4    & 512 & 16     & 0    & $0.02$ & $0.08$    \\
\hline
R14 & 30-40  & 128 & 0      & 0    &$\infty$& $\infty$  \\
R15 & 30-40  & 128 & 0      & 1    &$\infty$& $\infty$  \\
R16 & 30-40  & 128 & 8      & 0    & $0.29$ & $2.1$     \\
R17 & 30-40  & 128 & 16     & 0    & $0.15$ & $1.0$     \\
R18 & 30-40  & 128 & 16     & 1    & $0.15$ & $1.0$     \\
R19 & 30-40  & 128 & 32     & 0    & $0.07$ & $0.53$    \\
R20 & 30-40  & 128 & 64     & 0    & $0.04$ & $0.26$    \\
R21 & 30-40  & 128 & 128    & 0    & $0.02$ & $0.13$
\end{tabular}
\end{ruledtabular} \label{tab1}
\end{table}

\begin{figure}
\includegraphics[width=8.8cm]{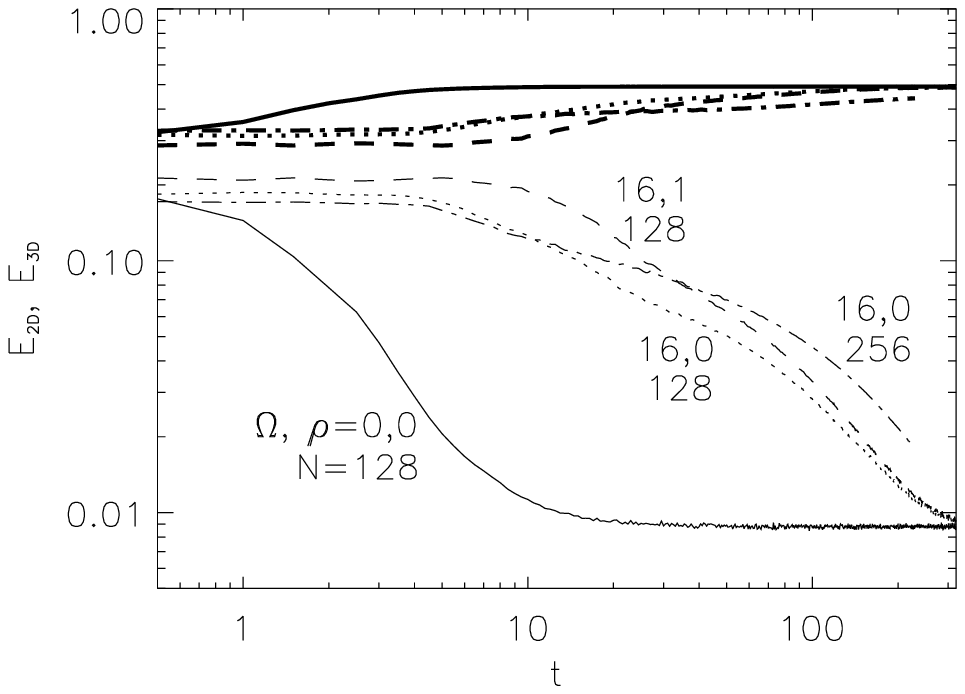}
\includegraphics[width=8.8cm]{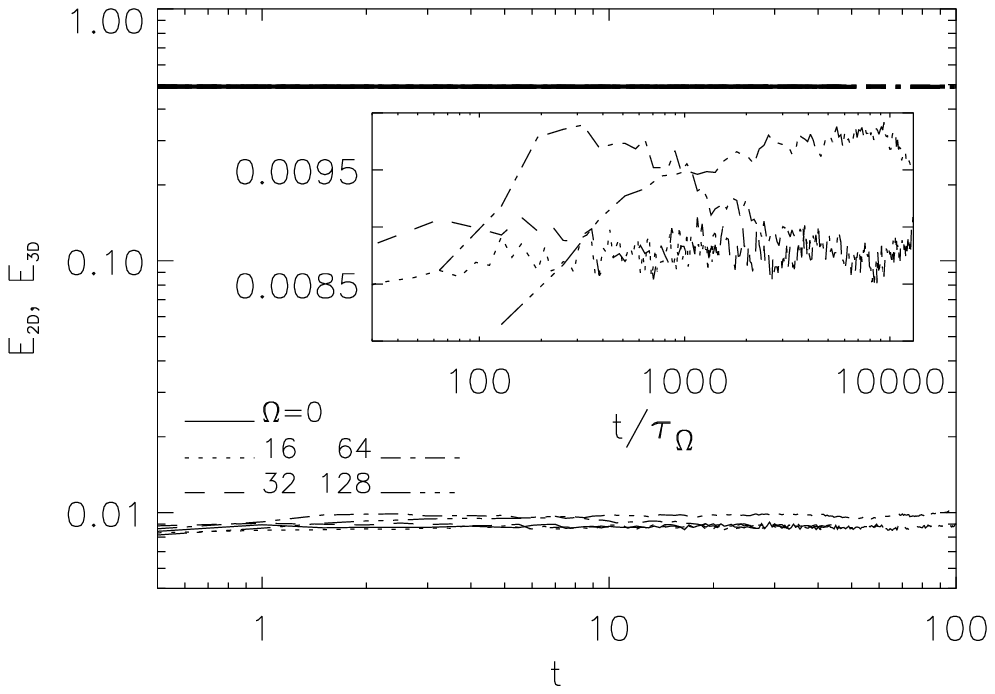}
\caption{{\it Top:} Energy $E_{3D}$ (thick) and $E_{2D}$ (thin) (systematically below $E_{3D}$) as a function of time in log-log coordinates for runs with initial conditions between $k=2$ and 4. Solid line: run R1, dots: run R2, dash-dots: run R11, all three with $\rho=0$  (see Table \ref{tab1} and labels on curves giving the pair of values $[\Omega,\ \rho]$) and the linear resolution $N$. Finally, with $\rho\approx 1$ we have run R3 (dash line). {\it Bottom:} $E_{3D}$ (thick) and $E_{2D}$ (thin) for runs with initial conditions between $k=30$ and 40. All runs have $\rho=0$ and $N=128$. Solid line: run R14, dots: R17, dash: R19, dash-dots: R20 and dash-triple-dot: R21. The inset shows $E_{2D}$ for runs R17, R19, R20, and R21 with time in units of the inertial period $\tau_\Omega$ (which goes from $\tau_\Omega \approx 0.03$ turnover times for run R17, to $\approx 0.004$ for run R21).}
\label{fig:ener}
\end{figure}

The Liouville theorem for fluid mechanics and magnetohydrodynamics (when coupling the fluid equation to the temporal evolution of the magnetic induction) was first derived by T.D. Lee \cite{tdlee} (see also \cite{rhk65}). One can find a simple derivation of the Liouville theorem for two-dimensional (2D) turbulent flows in \cite{rhk_montgo}. It expresses the incompressibility of the flow of dynamical variables in phase space (e.g., complex Fourier modes) using detailed conservation of quadratic invariants. Systems that can be described either by thermal equilibrium (zero-flux) or by non-equilibrium (finite flux) dynamics can also be found outside the realm of fluid mechanics. An example is provided by semi-conductor lasers described by the quantum Boltzmann equation \cite{lvov} for which, as in many cases of optical turbulence, both a direct and an inverse cascade can be observed in the presence of pumping and dissipation. Statistical equilibria have been derived for a variety of geophysical flows in the ideal case, in the presence of stratification \cite{waite04} or rotation \cite{bartello,bourouiba}, for shear flows \cite{Kaneda89}, and for quasi-geostrophic models \cite{Salmon76,Majda06}. An equivalent problem in magnetohydrodynamics, when an external magnetic field is imposed, was studied in \cite{shebalin}. One of the concerns of these studies is the existence of quasi-invariants linked to the quasi bi-dimensionalization of the flow. In such case, the possibility of their condensation at the gravest mode available to the truncated system may indicate the existence of an inverse cascade in the forced-dissipative case (see e.g., \cite{waite04}).

\begin{figure} \includegraphics[width=8.8cm]{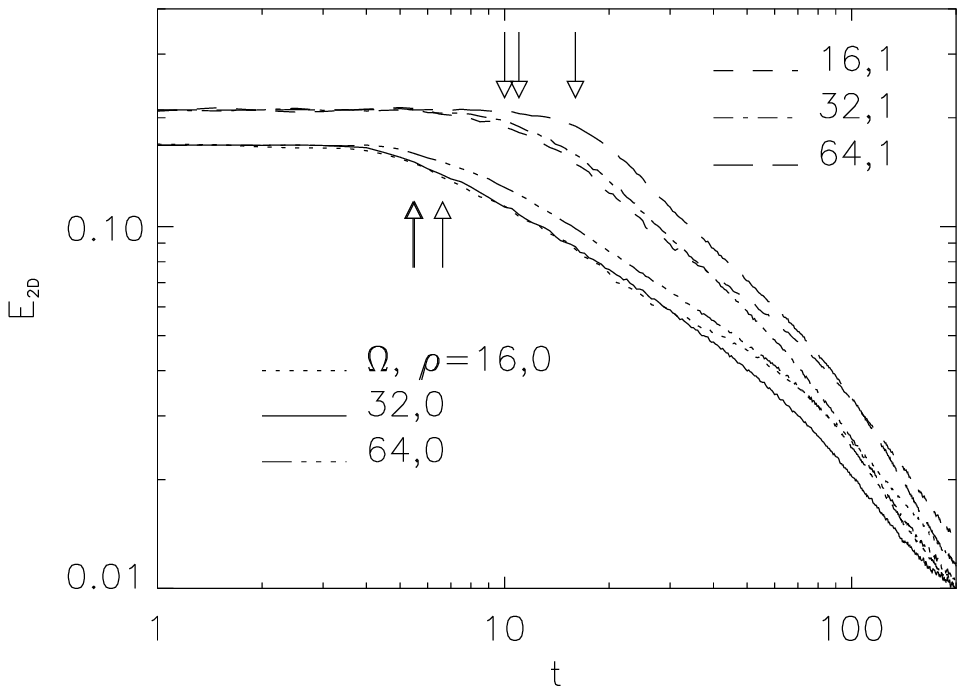} \includegraphics[width=8.8cm]{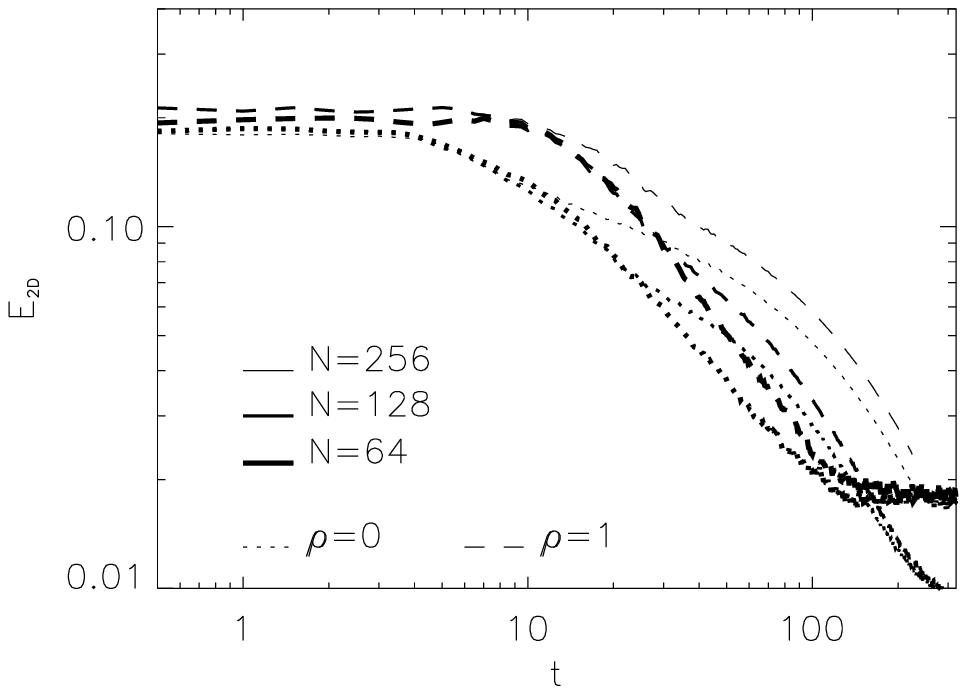}
\caption{{\it Top:} $E_{2D}$ in simulations with initial conditions between $k=2$ and 4, fixed spatial resolution ($N=128$), and increasing $\Omega$. Non-helical runs R2 ($\Omega=16$), R4 ($\Omega=32$), and R6 ($\Omega=64$), and helical runs R3 ($\Omega=16$), R5 ($\Omega=32$), and R7 ($\Omega=64$) are shown. The arrows indicate the time when $E_{2D}$ decreases below $90\%$ of its initial value, from left to right with increasing $\Omega$, above for helical runs and below for non-helical runs. The symbols are given in the figure, with $[\Omega,\ \rho]$ for $\rho=0$ in the bottom left, and for $\rho=1$ in the top right. {\it Bottom:} $E_{2D}$ in simulations with initial conditions between $k=2$ and 4, fixed rotation rate ($\Omega=16$), and increasing resolution (the higher, the thinner the line). Runs with zero helicity (dotted lines for runs R2, R8, and R11), and with net helicity (dashed lines for runs R3, R9, and R12) are shown.} 
\label{fig:helires}
\end{figure}

However, the late-time statistical equilibrium of many of these systems seems to be independent of the geophysical effects considered and only prescribed by the structure of the nonlinear terms in the Navier-Stokes equations. In addition, other invariants seem to play no significant roles. As an example, it was shown by Kraichnan \cite{rhk73} that in the presence of helicity, the global ideal equipartition equilibrium $E(k)\sim k^2$ is only slightly altered at small scale by the presence of helicity, while the large-scales show no sign of condensation. In this context, it is worth asking what happens in the rotating case, which can be viewed as a hybrid between the 2D and the three-dimensional (3D) cases. In the forced-dissipative case, the small scales are fully 3D and isotropy is recovered at small scale, whereas the large scales are dominated by an inverse cascade of energy. The ideal truncated problem was tackled recently in \cite{yamazaki} where it was shown that the establishment of the equilibrium is delayed in the presence of rotation in an anisotropic fashion. Furthermore, following analyses in \cite{newell,babin} and using a modal decomposition into wave, 2D, and 3D modes, \cite{bourouiba} found that in non-helical flows in the limit of strong rotation (small Rossby number Ro), decoupling between inertial waves and 2D coherent structures obtains until a time $t_{\ast}\sim \textrm{Ro}^{-2}$. After this time, coupling occurs but it can still be considered weak for long times and the system thus displays both 2D and 3D features. At this point, it is worth mentioning also that there is no consensus for the moment on whether perfect decoupling between the modes takes place for infinite rotation rate (see e.g., \cite{cambon2004}).

\begin{figure}
\includegraphics[width=8.8cm]{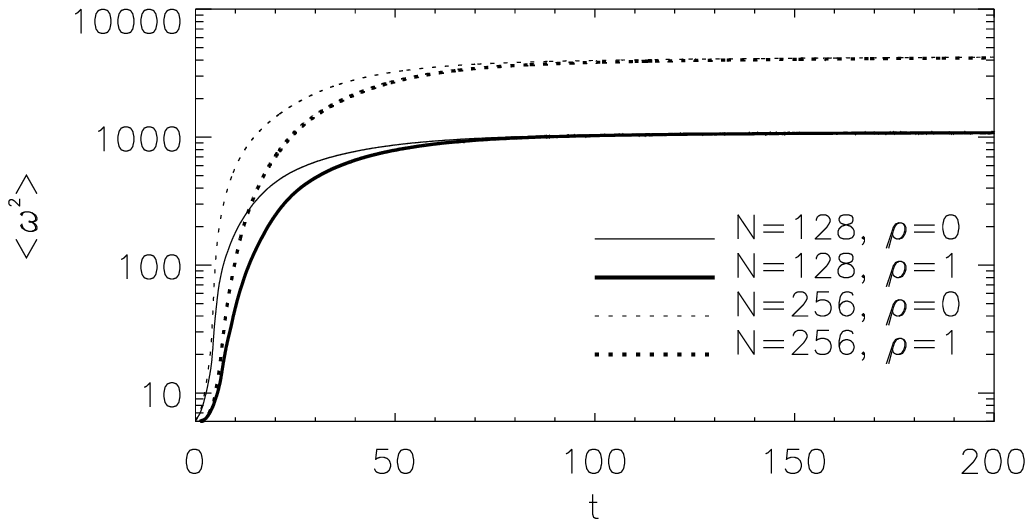}
\includegraphics[width=8.8cm]{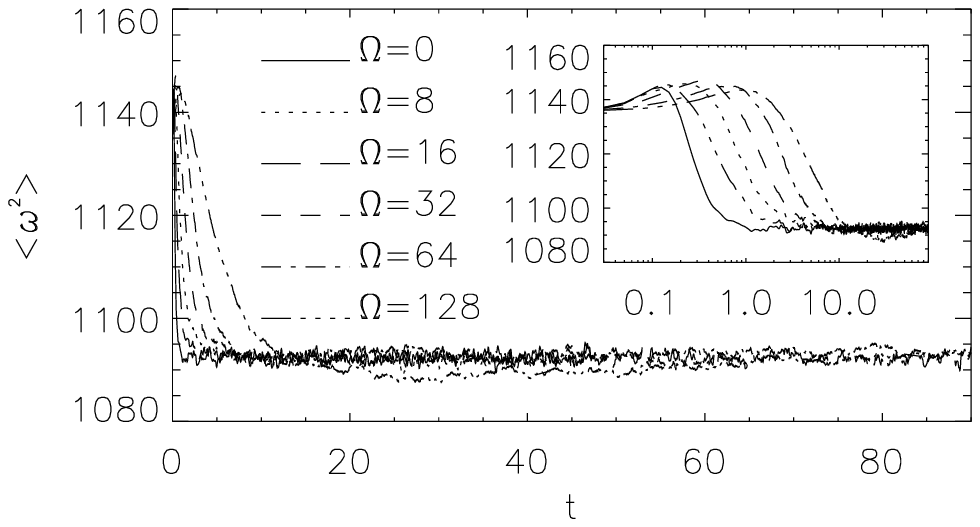}
\caption{Temporal evolution of the enstrophy. {\it Top:} Runs with $\Omega=16$, initial conditions between $k=2$ and 4, for different spatial resolutions and values of the relative helicity. {\it Bottom:} Runs with initial conditions between $k=30$ and 40, $N=128$, $\rho=0$, and different values of $\Omega$. The inset shows the same curves with the time in logarithmic scale.}
\label{fig:ens_res}
\end{figure}

We revisit here some of these results and extend our analysis to the helical case and to the dynamics of intermediate times during which an energy (and helicity) distribution can be observed reminiscent of  dissipative dynamics, as shown in the non rotating case in \cite{meb,meb_H}. In the next section, we describe briefly our procedure and move on in Sect.~\ref{s:global} to discuss some of the predictions for early times in these systems, before thermalization is reached. Then, in Sect.~\ref{ss:results} we examine non-helical and helical flows initialized with perturbations either in the small or in the large scales. Section \ref{s:conclu} presents our conclusions.

\section{The procedure} \label{s:proc}
We consider the equations for an incompressible and inviscid three-dimensional rotating fluid for the velocity field ${\bf u}$, with $\nabla \cdot {\bf u} =0$. In the rotating frame, they read:
\begin{equation}
\frac{\partial {\bf u}}{\partial t} + \mbox{\boldmath $\omega$} \times
    {\bf u} + 2 \mbox{\boldmath $\Omega$} \times {\bf u}  =
    - \nabla {\cal P} \ .
\label{eq:momentum}
\end{equation}
Here $\mbox{\boldmath $\omega$} = \nabla \times {\bf u}$ is the vorticity and ${\cal P}$ is the total pressure  modified by the centrifugal term, which is obtained self-consistently by taking the divergence of Eq.~(\ref{eq:momentum}), using the incompressibility condition, and solving for the resulting Poisson equation. We use a pseudo-spectral method in a $(2\pi)^3$ box with periodic boundary conditions. The code is fully parallelized using the message passing (MPI) library \cite{Gomez05a,Gomez05b}; spatial resolutions in this paper are up to $N^3=512^3$ with $N$ points in each direction. The temporal scheme is a fourth order Runge-Kutta and the code uses the $2/3$ dealiasing rule, resulting in a truncated set of modes from $k_\textrm{min}=1$ to a maximum wave number $k_\textrm{max} = N/3$. The total energy $E=\left<|{\bf u}|^2 \right>/2$ and the total helicity $H=\left<{\bf u} \cdot \mbox{\boldmath $\omega$} \right>$ are inviscid invariants preserved by the truncation \cite{Moffatt69, rhk73}, with respective isotropic Fourier spectra $E(k)$ and $H(k)$; note that the Schwartz inequality implies $|H(k)|/kE(k) \le 1$. The relative helicity will then be defined in the following as 
\begin{equation}
\rho = \int{H(k) dk}\Bigg/ \Bigg(\int{k E(k) dk} \, \Bigg) .
\end{equation}

\begin{figure}
\includegraphics[width=8.8cm]{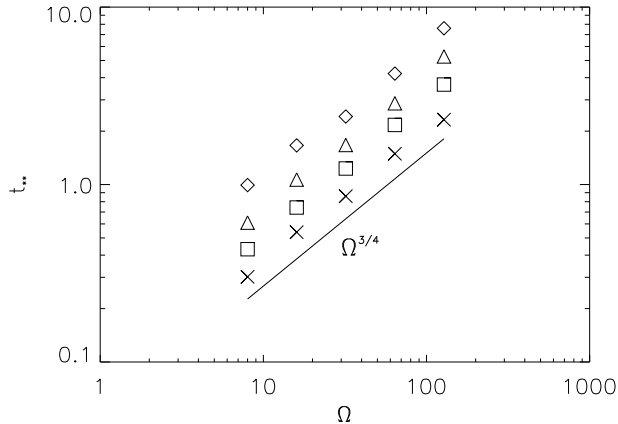}
\caption{Scaling of the time $t_{\ast\ast}$ for a given scale to reach thermalization as a function of the imposed rotation; $t_{\ast\ast}\sim \Omega^{3/4}$ appears as a good fit. Different symbols correspond to different choices of the thermalization level (see text for details). Runs have initial conditions between $k=2$ and 4, $N=128$, and $\rho=0$.}
\label{fig:scaling}
\end{figure}

The rotation axis is taken to be in the $z$ direction with $\mbox{\boldmath $\Omega$} = \Omega \hat{z}$, $\Omega$ being the rotation frequency. Finally, the Rossby number is defined as usual as:
\begin{equation}
\textrm{Ro} = \frac{U_{rms}}{2 \Omega L_0}  \ .
\end{equation}
$U_\textrm{rms}$ is the r.m.s.~velocity taken to be of order unity, and $L_0=2\pi/k_0$ is the characteristic scale of the initial conditions ($k_0$ in this expression corresponds to the minimum wavenumber with non-zero amplitude in the initial conditions). One can also define the micro Rossby number based on the r.m.s.~vorticity
\begin{equation}
Ro_\omega=\frac{\omega_\textrm{rms}}{2\Omega} \ ,
\end{equation}
measuring the intensity of the small-scale vorticity with respect to the imposed rotation.

In order to examine long-time properties of the inviscid flow, computations were run for over $100$ (in some cases up to $640$) turnover times $\tau = L/U_\textrm{rms}$ in terms of a unit length $L$ (in the following, time $t$ in all figures is in units of $\tau$ unless explicitly stated). The turnover time at the scale of the initial conditions is then given by $\tau_0 = L_0/U_\textrm{rms} = (L_0/L) \tau$ and the inertial period is $\tau_\Omega=\textrm{Ro} \tau_0$. 

Two types of initial conditions are taken, one in the small scales, with the initial velocity field defined in the band of wavenumbers $k_0 \in [30,40]$, and one in the large scales, with $k_0 \in [2,4]$. The former case is to ensure that the effect of helicity is observable in the absolute equilibrium derived by Kraichnan \cite{rhk73}, whereas the latter is to examine a sufficiently resolved energy spectrum in the direct transfer to small scales. Also, the two sets of initial conditions allow us to study the dependence with scale of the time to reach the statistical equilibrium. The initial amount of helicity is controlled using the method given in \cite{patterson}, coupling two random fields by imposing a prescribed angle (in physical space) between them. Some of the properties of the runs described in this paper are summarized in Table \ref{tab1}.

Assuming a truncated system of modes and considering the two quadratic invariants, the energy $E=\int E(k)dk$ and the helicity $H=\int H(k)dk$, the macro-canonical probability density function is proportional to $C\exp{[-\alpha E -\beta H]}$. This results in the ideal solutions for the spectra \cite{rhk73}:
\begin{equation}
 E(k) = \frac{2\alpha k^2}{D(k)} \ \ \ ,  \ \ \  H(k) = \frac{2\beta k^4}{D(k)}
\label{eq:equil} \end{equation}
with
$$D(k) = \alpha^2 - \beta^2 k^2 \ .$$
In these solutions, $\alpha$ and $\beta$ can be seen as temperatures associated with the two invariants $E$ and $H$, constrained by $\alpha >0$ and $|\beta|  k_{max}< \alpha$ (for given initial energy and helicity, these equations can be solved for $\alpha$ and $\beta$; see \cite{meb_H}). Note that helicity and thus $\beta$ are not positive definite: they are pseudo-scalars, changing sign when one goes from a right-handed to a left-handed frame of reference. Since $H(k)/kE(k) \sim k$ in Eq. (\ref{eq:equil}), eddies at a smaller scale are more helical, contrary to the dissipative case for which this ratio is observed to vary as $1/k$ in the non rotating case, thereby indicating a slow $\sim k^{-1}$ recovery of mirror-symmetry as $k \rightarrow \infty$. This discrepancy in the relative behavior of spectra may be the origin of persistence of helical motions in the dissipative range of viscous flows \cite{Mininni06,meb_H}. With $\beta\equiv 0$ when $H\equiv 0$, one recovers in the ideal case the equipartition of energy between Fourier modes, with a number of modes per spherical shell of radius $k$ and unit width proportional to $k^2$. In the presence of solid-body rotation the Coriolis force, being conservative and linear in the velocity, does not alter the long-time statistical equilibria given in equation (\ref{eq:equil}), and thus the isotropic distribution of energy is also unaltered.

The numerical conservation of energy and helicity is therefore essential when performing ideal truncated runs and thus we have systematically monitored their evolutions. For example, in runs with $\Omega=128$, we had to resort to a time-step 32 times smaller than in runs with $\Omega=0$ (resulting, e.g., in $\Delta t =5\times 10^{-3}$ for the former and $\Delta t \approx 1.5\times 10^{-4}$ for the latter when $N=128$). In such a case, the energy is conserved to within $0.5\%$ after $t\approx 100$ for $\Omega=128$, and better still for smaller rotation rates at the same resolution (e.g., at $0.02\%$ for $\Omega=64$ at the same time).
\begin{figure}
\includegraphics[width=8.8cm]{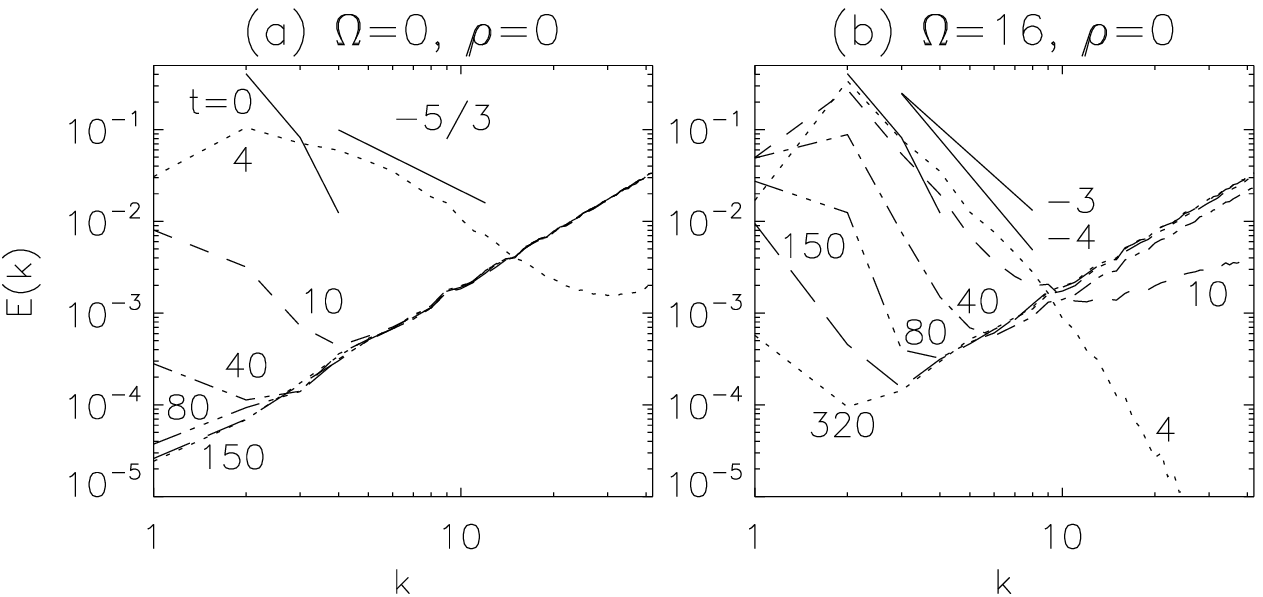}
\includegraphics[width=8.8cm]{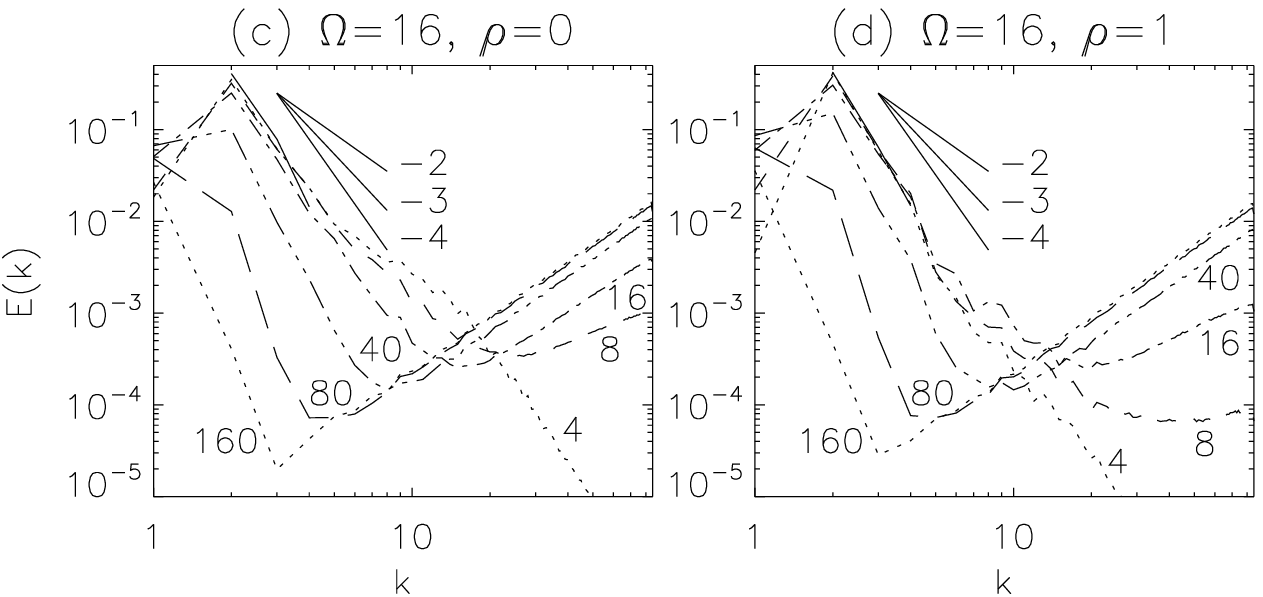}
\caption{Time evolution of the isotropic energy spectrum. Above are runs with $N=128$: (a) R1 with $\Omega=0$ and $\rho=0$, and (b) R2 with $\Omega=16$ and $\rho=0$; in (a) and (b) times are $t=0$ (solid), $t=4$ (dotted), $t=10$ (dashed), $t=40$ (dash-dotted), $t=80$ (dash-triple-dotted), $t=150$ (long dashed), and $t=320$ (dotted). Below are runs with $N=256$: (c) R11 with $\Omega=16$ and $\rho=0$, and (d) R12 with $\Omega=16$, and $\rho\approx 1$; in (c) and (d) times are (with the same line style order as above) $t=0$, $t=4$, $t=8$, $t=16$, $t=40$, $t=80$, and $t=160$. Different slopes are indicated only as references. All runs build a $k^2$ tail over time.}
\label{fig:en_evol}
\end{figure}

\begin{figure}
\includegraphics[width=8.8cm]{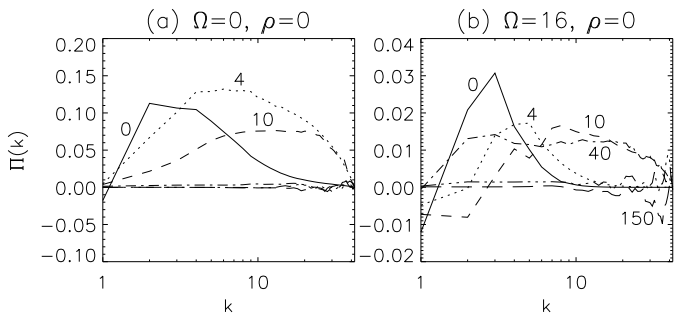}
\caption{Energy flux $\Pi(k)$ in runs R1 (a) and R2 (b). Symbols for different times are as in Fig.~\ref{fig:en_evol}(a-b).}
\label{fig:flux}
\end{figure}

\section{Is there a large-scale ideal condensate in the rotating case?} \label{s:global}
As mentioned previously, an inverse energy cascade has been observed in several numerical simulations of driven dissipative rotating turbulence whereas the (isotropic) statistical equilibria show no sign of large-scale condensation. We will show that in our simulations, the obtained equilibria are independent of the resolution used. However, very long integrations are needed in the presence of rotation and helicity to reach these final thermalized states. As a result, we next resort to a series of direct numerical simulations (DNS) at moderate resolution and computing for long times to examine further this problem.

It is useful to decompose the total velocity field in the following manner \cite{Waleffe93, babin, bourouiba_bartello}. We define a preferred direction $\hat {\bf z}$, parallel to the rotation axis. A general three-dimensional flow ${\bf u}$ can then be written as the sum of a strictly two-dimensional velocity ${\bf u}_{\perp}(k_{\perp},k_{\parallel}= 0)$ of energy $E_\perp$, a vertical component $u_z(k_{\perp},k_{\parallel}= 0)$ of energy $E_{z}$, and the remaining three-dimensional component ${\bf u}_3(k_{\perp},k_{\parallel}\not= 0)$ of energy $E_\textrm{3D}$. The sub-indices $\perp$ and $\parallel$ refer to the rotation axis. The total kinetic energy is:
\begin{eqnarray}
E &=& \frac{1}{2} \int |{\bf u}({\bf x})|^2 d^3{\bf x} = \int E(k) dk = \nonumber \\
&=& E_\textrm{3D}+E_\perp+E_z = E_\textrm{3D}+E_\textrm{2D} ,
\label{eq:decomp} \end{eqnarray}
where $E_\textrm{2D}=E_\perp+E_z$ is the total energy in the modes with $k_\parallel=0$. Note that there is a one to one correspondence \cite{bourouiba_bartello} between $E_\textrm{3D}$ and $e(k_\perp,k_\parallel \ne 0)$, $E_\perp \propto [e -Z_\textrm{PA}](k_\perp, k_{\parallel}=0)$, and $E_z\propto [e +Z_\textrm{PA}](k_{\perp}, k_{\parallel}=0)$, where $e(k_\perp,k_\parallel)$ is the two-dimensional axisymmetric energy spectrum, and $Z_\textrm{PA}(k_\perp,k_\parallel)$ is the polarization anisotropy. The spectra can be expressed in Cartesian coordinates $(k_\perp,k_\parallel)$, or in polar coordinates $(k,\cos \theta)$ where $\theta$ is the angle between the vector ${\bf k}$ and the vertical axis. These scalar spectra $e$ and $Z_\textrm{PA}$, together with the helicity spectrum, are the dynamical variables used in studies of rotating turbulence using spectral closures \cite{Cambon89}.

\begin{figure}
\includegraphics[width=8.8cm]{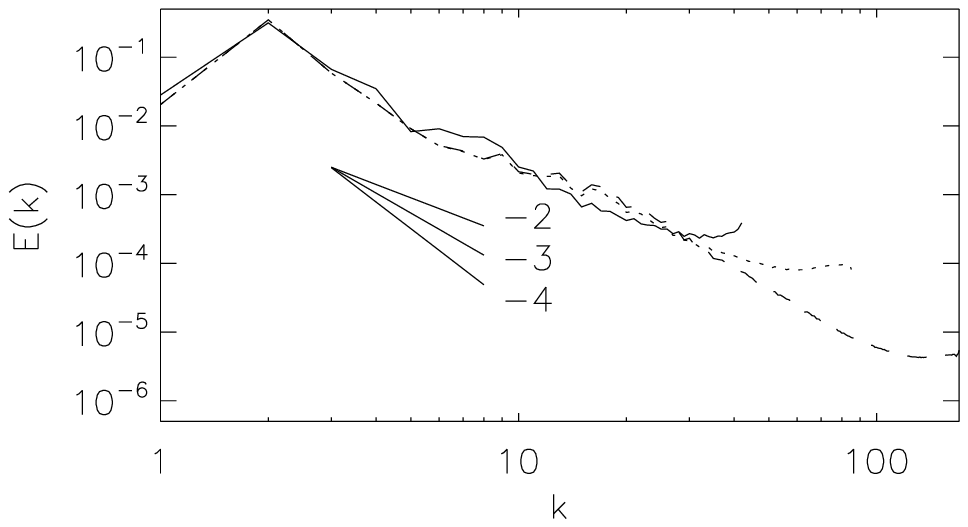}
\includegraphics[width=8.8cm]{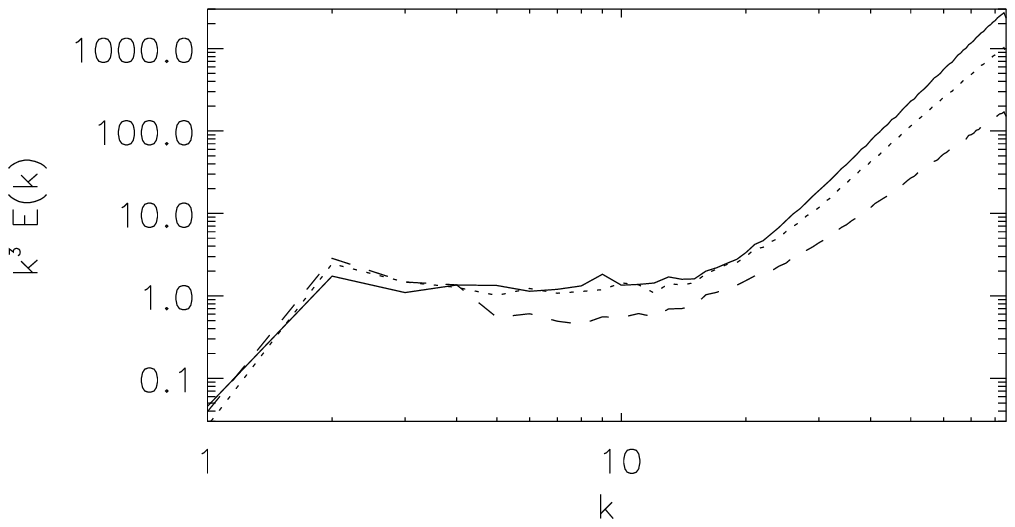}
\caption{{\it Top:} Isotropic energy spectrum at $t=5$ (at the beginning of the intermediate phase when thermalization begins at small scale) in runs without helicity and with $\Omega=16$: run R2 (solid), R11 (dotted), and R13 (dashed), with increasing resolution from $N=128$ to $N=512$. {\it Bottom}: energy spectrum compensated by $k^{-3}$ at $t=10$ in $256^3$ runs R10 (solid \ADD{line}, $\Omega=8$ and $\rho=0$), R11 (dotted, $\Omega=16$ and $\rho=0$), and R12 (dashed, $\Omega=16$ and $\rho = 1$).}
\label{fig:spec_comp}
\end{figure}

Bourouiba \cite{bourouiba} argues, by considering only resonant triads, that for small enough Rossby number, the 2D and 3D modes are effectively decoupled for times $t<t_{\ast}$. This implies separate conservation properties of $E_\textrm{3D}$, $E_\perp$, and $E_z$. In that case, a different statistical equilibrium to the one described in Sect. \ref{s:proc} can be derived, which leads to separate predictions for the horizontal dynamics of the flow. For late times ($t>t_*$), the isotropic statistical equilibrium is recovered. 

A non-ideal example where decoupling between 2D and the 3D modes can be observed may be given by freely decaying rotating flows. In \cite{tomas_trieste}, it was found that in this case the 2D and 3D energies decay at different (and independent) rates. If we follow the estimation of coupling being effective after the time $t_{\ast}\sim \textrm{Ro}^{-2}$ \cite{bourouiba_bartello}, and taking into account that the Rossby number itself decays as $t^{1/2}$ according to the study in \cite{tomas_trieste}, we conclude that $t_{\ast}\sim t$. In other words, $t_{\ast}$ may never be reached. It is not clear, however, that in the forced case where the Rossby number is maintained at a constant value, such a decoupling can remain permanently \cite{Chen05}.

\section{Numerical results}\label{ss:results}
\subsection{Temporal dynamics properties} \label{ss:temp}

We first consider an initial condition concentrated in the large scales (runs R1 to R13). An isotropic excitation with random phases and fast decaying energy spectrum is put at $t=0$ between $k=2$ and 4 in Fourier space, such that $U_{rms} = 1$ and $\tau_0 \approx 3.1$. The temporal evolution of the energies, specifically $E_{3D}$ and $E_{2D}$ (see Eq.~(\ref{eq:decomp})) are displayed in Fig.~\ref{fig:ener} for rotation rates $\Omega=0$ and $\Omega=16$, and for different spatial resolutions ranging from $N=128$ to $256$. Simulations with and without helicity are considered.

In the $128^3$ run with $\Omega=0$ and $\rho=0$, $E_\textrm{3D}$ grows fast to $\approx 0.49$ and $E_\textrm{2D}$ decays to $\approx 0.01$ after $t\approx 10$. These values correspond to the isotropic statistical equilibrium, where all the modes have thermalized and have the same energy per mode ($E_\textrm{3D}$ is larger as there are more 3D modes than modes in the $k_\parallel=0$ plane). In the $128^3$ run with $\Omega=16$, these thermalized values are only reached after $t \approx 300$. In the presence of rotation not only the decay towards the isotropic statistical equilibrium is slowed down; at early times (up to $t\approx 4 \approx 1.3\tau_0$) a transient regime develops for which $E_\textrm{2D}$ and $E_\textrm{3D}$ seem to be independently conserved. Details of this regime for different Rossby numbers can be found in \cite{bourouiba}. Here we consider the effect of resolution and helicity in the time when the quasi-conservation breaks down. To first illustrate this, a $256^3$ run with $\Omega=16$ is also shown in Fig.~\ref{fig:ener}. The change in resolution does not seem to affect the quasi-conservation of $E_\textrm{2D}$ and $E_\textrm{3D}$, although the following decay towards the isotropic statistical equilibrium takes place at a slower pace at larger resolution. A similar result was obtained for a $512^3$ simulation (not shown). Finally, the presence of helicity in a $128^3$ run with $\Omega=16$ increases (by close to a factor 2) the time during which $E_\textrm{2D}$ and $E_\textrm{3D}$ can be considered as quasi-invariants.

The effect of resolution and helicity in the quasi-conservation can be further explored considering runs R2 to R12. Runs R2 to R7 span $\Omega$ from 16 to 64 with initial relative helicity $\rho$ of 0 or 1 at fixed resolution ($N=128$). The time evolution of $E_{2D}$ in these runs is shown in Fig.~\ref{fig:helires}. As $\Omega$ is increased, the time during which $E_{2D}$ (and $E_{3D}$) can be considered quasi-invariant increases, and the differences between the non-helical and helical cases become more evident. In all cases, helicity extends the time of quasi-conservation. At the end of the quasi-conservation phase, $E_{2D}$ in all runs decreases with nearly the same rate independently of $\Omega$, and depending only on whether the runs have helicity or not.

Figure \ref{fig:helires} also shows the effect of changing resolution. Runs with fixed $\Omega$ are shown with and without helicity, while changing $N$ from 64 to 256. The time during which $E_{2D}$ is quasi-invariant seems insensitive to the linear resolution, while the rate of change after the quasi-conservation (as well as the final value reached at late times) depends on $N$. As a result, high resolution is not necessary to study the early time (or, as shown next, the very late time) dynamics of the ideal truncated systems. Simulations at higher resolution in Table \ref{tab1} will mostly be useful to consider the spectrum that develops at intermediate scales during the transition from the quasi-invariant regime to the isotropic thermalized regime.

\begin{figure}
\includegraphics[width=8.8cm]{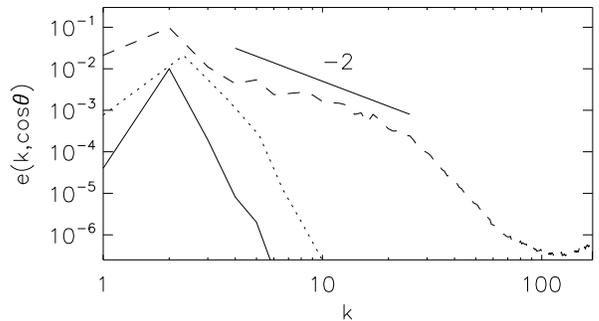}
\caption{Two dimensional energy spectrum $e(k,\cos \theta)$ in run R13 at $t=10$, for $\theta=0$ (dashed), $\theta=\pi/4$ (dotted), and $\theta=\pi/2$ (solid). A slope of $-2$ is shown as a reference.}
\label{fig:2Dspec}
\end{figure}

The case of initial conditions concentrated in the small scales (runs R14 to R21) is also shown in Fig.~\ref{fig:ener}. Again an isotropic excitation at $t=0$ is considered, but now between wavenumbers $k=30$ and 40, with $U_{rms} = 1$ and $\tau_0 \approx 0.2$. Except for very early times, the $E_\textrm{2D}$ and $E_\textrm{3D}$ energies remain constant and close to $\approx 0.49$ and $\approx 0.01$ respectively. However, unlike the previous runs and the cases considered in \cite{bourouiba}, the constancy of $E_\textrm{2D}$ and $E_\textrm{3D}$ should not be interpreted as a conservation of quasi-invariants (note it also takes place for $\Omega=0$). With the initial conditions at small scales, the system goes faster to the isotropic thermalized equilibrium (the turnover time is shorter in this case), and the constancy of $E_\textrm{2D}$ and $E_\textrm{3D}$ is simply the result of all modes having thermalized with the same energy. Once the thermal equilibrium is achieved, the solution is flux-less and thus no exchange of energy takes place. An approximate conservation of the quasi-invariants may happen at much earlier times, as shown in the inset of Fig.~\ref{fig:ener}. Runs R20 with $\Omega=64$ and R21 with $\Omega=128$ show an early increase of $E_\textrm{2D}$ followed by approximate constancy, to later decay to the isotropic thermalized value.

As a result, the time to reach equilibrium is delayed in the presence of either rotation or helicity, but the evolution of such flows drives them to the three-dimensional equilibrium given in Eq.~(\ref{eq:equil}) as will be shown in more detail later. This delay can also be observed in the time evolution of the enstrophy. In Fig.~\ref{fig:ens_res} we contrast the effects on its temporal evolution of the scale of the initial conditions, of the initial relative helicity, and of the spatial resolution. In the runs with initial conditions between $k=2$ and 4, $\left< \omega^2 \right>$ grows fast and reaches its saturated value (which corresponds to the isotropic thermalized state) after $t\approx 100$. Increasing the resolution does not affect this time, although it substantially increases the saturation value of the mean square vorticity (as more wavenumbers are available for the thermalization). On the other hand, the presence of helicity slows down the decay towards the thermalized state. Similar results are obtained in the runs with initial conditions between $k=30$ and 40. In this case, $\left< \omega^2 \right>$ decreases with time as the initial enstrophy, concentrated at small scales, is distributed among all wavenumbers as the system thermalizes. These runs also illustrate how the time to reach the thermalization increases as $\Omega$ is increased.

Defining an intermediate wavenumber between $k_\textrm{min}$ and $k_\textrm{max}$, we can compute the time $t_{\ast\ast}$ at which thermalization has reached that wavenumber, in the sense that its amplitude does not evolve except for thermal fluctuations around the mean after that time. In practice, this is equivalent to measuring the time at which a certain value of $\left< \omega^2 \right>$ is reached in Fig.~\ref{fig:ens_res}. Fig.~\ref{fig:scaling} displays the scaling of that time as a function of imposed rotation rate; a $\Omega^{3/4}$ range appears clearly. We have checked that this scaling is insensitive to the choice of scale at which it is computed. The different symbols in the figure correspond to different choices of thermalization level, with diamonds corresponding to $\left< \omega^2 \right>$ within $1\%$ of the thermalized value, triangles to $2\%$, squares to $3\%$, and crosses to $4\%$. Note this time should not be confused with the time $t_{\ast}$ for coupling of 2D and 3D modes that results in the end of the conservation of quasi-invariants.

\begin{figure}
\includegraphics[width=8.8cm]{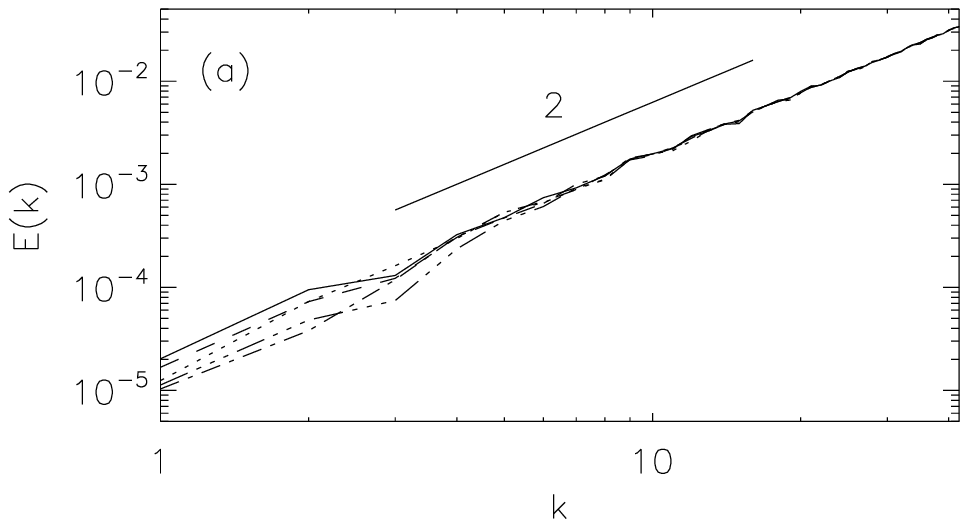}
\includegraphics[width=8.8cm]{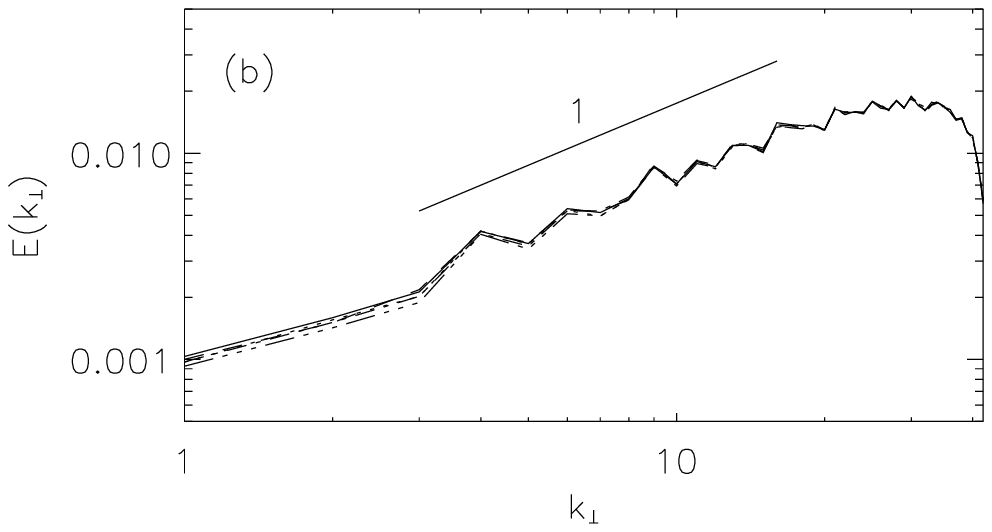}
\includegraphics[width=8.8cm]{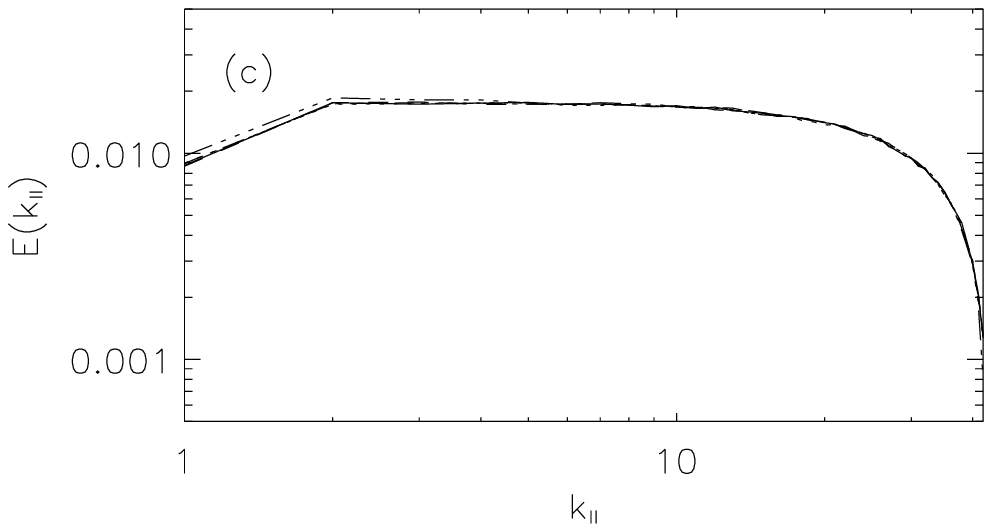}
\caption{(a) Isotropic energy spectrum, (b) reduced perpendicular spectrum, and (c) reduced parallel spectrum at $t=44$ in non-helical runs R14 (solid, $\Omega=0$),  R17 (dotted, $\Omega=16$), R19 (dashed, $\Omega=32$), R20 (dash-dotted, $\Omega=64$), and R21 (dash-triple-dotted, $\Omega=128$). Slopes are given as references.}
\label{fig:nohel}
\end{figure}

\begin{figure}
\includegraphics[width=8.8cm]{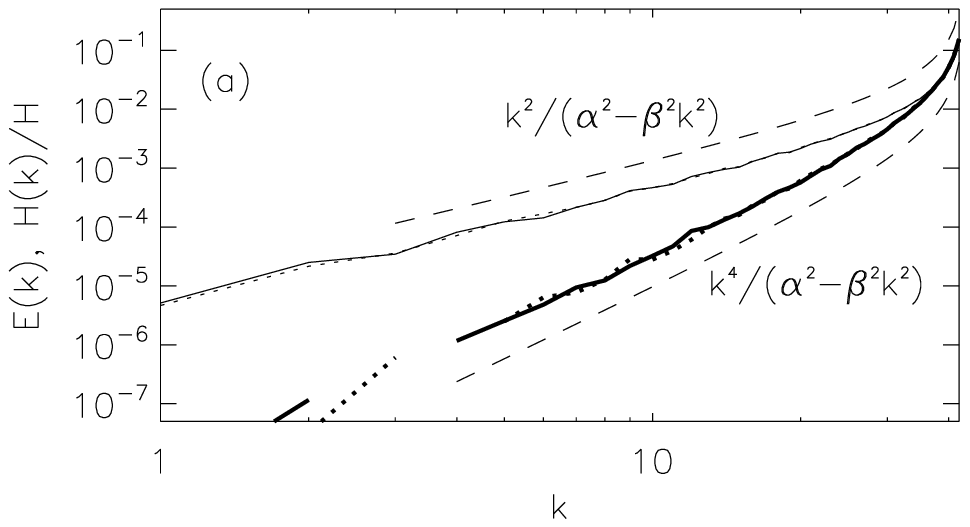}
\includegraphics[width=8.8cm]{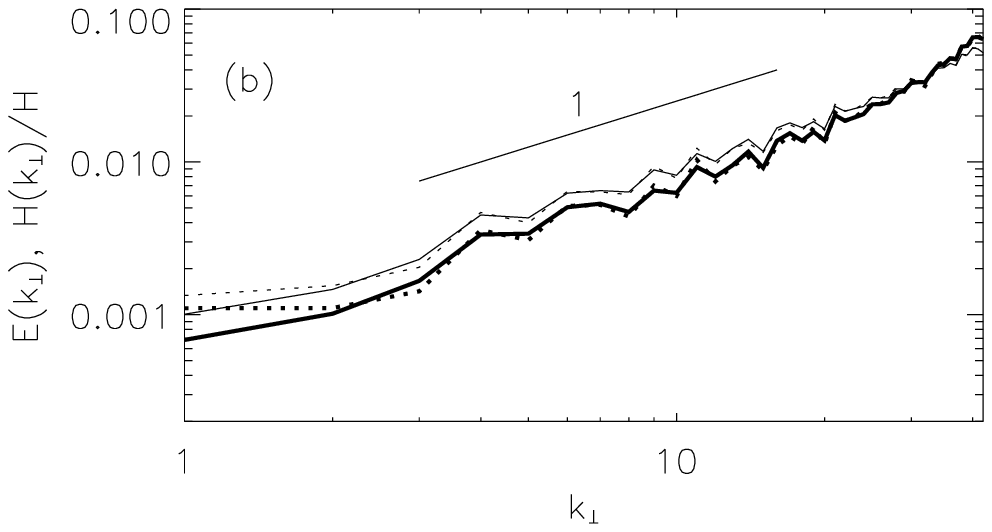}
\includegraphics[width=8.8cm]{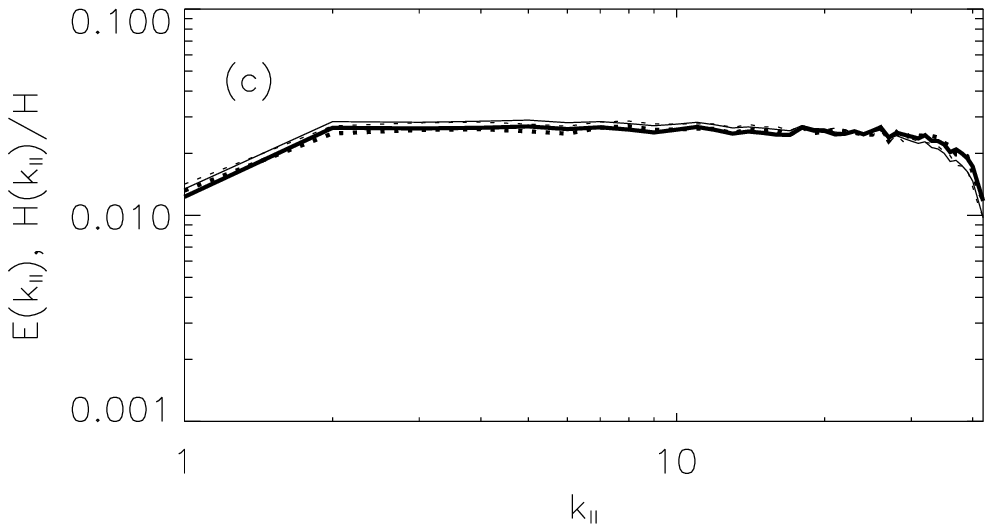}
\caption{(a) Isotropic spectra, (b) reduced perpendicular spectra, and (c) reduced parallel spectra of the energy (thin) and of the helicity (thick) at $t=55$ in helical runs R15 (solid, $\Omega=0$) and R18 (dotted, $\Omega=16$). Slopes are given as references. Helicity spectra are normalized by the total helicity. In (a), the isotropic prediction of Kraichnan \cite{rhk73} is shown in dashed lines, with values of $\alpha$ and $\beta$ obtained from solving Eq. (\ref{eq:equil}).}
\label{fig:hel}
\end{figure}

\subsection{Large-scale dynamics and early time evolution} \label{ss:large}

As it will be confirmed in the next subsection, the ideal flow ends up in a state of equipartition of energy as predicted by T.D. Lee \cite{tdlee} (or its generalization to helical flows \cite{rhk73}). However, the path to get there is interesting as it uncovers a ``dissipative'' transient that displays, in the non-rotating case, a Kolmogorov spectrum. Although the total energy is conserved, the large wavenumber $k^2$ tail -- established at intermediate times as a partial thermalization -- plays the role of a turbulent viscosity for the large scales, which can undergo a standard turbulence dynamics. This was shown both for the non-helical case \cite{meb} and for the helical case \cite{meb_H}. We thus examine this temporal behavior in the presence of rotation, by considering the runs with initial conditions at large scales in the band of wavenumbers between $k=2$ and 4. Note that this corresponds to the transition of the system from the quasi-invariant regime to the isotropic statistical equilibrium.

The time evolution of the isotropic energy spectrum is shown in Fig.~\ref{fig:en_evol} for a non-helical non-rotating run (run R1), two rotating non-helical runs (R2 and R11, at different resolutions), and a rotating helical run (R12). Several times are displayed in order to show the progressive advance of thermalization towards large scales and the dynamics at large scale. Run R1 at early times develops a spectrum compatible with a $\sim k^{-5/3}$ law, and as the thermalized range widens the spectrum goes progressively to the $\sim k^2$ solution. For non-rotating flows, this behavior was studied before (we refer the reader for details to Refs. \cite{meb,Frisch08,meb_H,Wan09}). In the rotating case (Run R2), a dissipative-like transient also develops as the system transits from the regime with quasi-conserved $E_\textrm{2D}$ and $E_\textrm{3D}$ to the isotropic thermalized equilibrium, but the large-scale spectrum is steeper in this case. As will be shown next, the steeper spectrum seems to be consistent with a $\sim k^{-3}$ law. As in the non-rotating case, the small scales evolve towards the isotropic $\sim k^2$ solution, but the time for the large-scales to reach thermalization is substantially increased in the presence of rotation. Changing spatial resolution does not seem to affect these results, but the presence of helicity makes the large-scale spectrum even steeper (closer to $\sim k^{-4}$, compare (c) and (d)).

Figure \ref{fig:flux} gives the energy fluxes at different times for the $128^3$ non-helical runs with $\Omega=0$ and 16 (runs R1 and R2). These fluxes evolve towards zero as time elapses, corresponding to zero-flux statistical equilibria solutions, but at intermediate times they show clear energy exchanges between Fourier modes. In the case with $\Omega=0$, a range with positive flux (which can be associated to the $\sim k^{-5/3}$ range) is observed before the flux goes to zero at all scales. In the presence of rotation, besides a range with approximately constant direct flux, a change of sign of the flux at large scales is observed which indicates a tendency toward a large-scale condensation, associated to the inverse cascade of energy in the forced-dissipative case. However, the final state lacks such condensate as already stated, and the system evolves towards zero-flux equilibria (although at later times than in the non-rotating case).

The effect of resolution and of helicity in the large-scale energy spectrum is further illustrated in Fig.~\ref{fig:spec_comp}, which shows the spectrum at $t=5$ in runs with $\Omega=16$ and $\rho = 0$ for increasing spatial resolution up to $N=512$. At this early time little thermalization $\sim k^2$ is present yet. All runs show spectra compatible with a $k^{-3}$ law at large scales although a knee followed by a slightly shallower spectrum is observed in the higher-resolved runs. As will be shown next, this knee in the spectrum is the result of anisotropies developing in the flow. Figure \ref{fig:spec_comp} also gives the energy spectrum compensated by $k^{-3}$ at $t=10$ in runs with $256^3$ grid points and for various rates  of rotation and of relative helicity. The change of $\Omega$ does not affect the power law, although in the presence of helicity a steeper energy spectrum obtains, in agreement with the results shown in Fig.~\ref{fig:en_evol}. Also, a slightly shallower spectrum than $k^2$ obtains in the smallest scales, indicating again a slowing-down of the dynamics.

Both in the non-helical and helical cases in the presence of rotation, a spectrum different from Kolmogorov is obtained, with an inertial index between $-3$ and $-4$. Such spectra also differ from the ones found in driven dissipative studies of rotating turbulence, namely $E(k)\sim k^{-2}$ or a bit steeper in the helical case. For non-helical rotating turbulence, a $k^{-3}$ spectrum has been advocated using a spectral closure \cite{cambon2004} because of the distribution of energy for different angles between the wavevector ${\bf k}$ and the rotation axis, the classical $k^{-2}$ spectrum being found only for modes with $k_\parallel \approx 0$. Indeed the present simulations confirm this prediction. The axisymmetric two-dimensional energy spectrum $e(k,\cos \theta)$ \cite{Cambon97,Bellet06} is shown in Fig.~\ref{fig:2Dspec} for different values of $\theta$ for run R13 at $t=5$. The differences in the spectra for different directions indicate the development of anisotropies, with most of the energy in modes with $k_\parallel = 0$ ($\theta=0$) and following $e(k,\cos \theta=1) \sim k^{-2}$. As $\theta$ increases towards $\theta = \pi/2$ (corresponding to wave vectors ${\bf k} = k_\parallel \hat{z}$) the energy content decreases and the spectrum becomes steeper. Note that an isotropic $k^{-2}$ spectrum can be recovered using a phenomenological analysis taking into account the slowing down of nonlinear transfer due to inertial waves, still assuming isotropy \cite{dubrulle, zhou}. The extension of this argument to the anisotropic case is straightforward and leads to $E(k_{\perp}, k_{\parallel})\sim k_{\perp}^{-5/2} k_{\parallel}^{-1/2}$ \cite{mangeney}. However, the anisotropic spectra observed here in the transient phase are steeper and are more likely attributed to an angular dependence as advocated in \cite{cambon2004}.

\begin{figure}
\includegraphics[width=8.8cm]{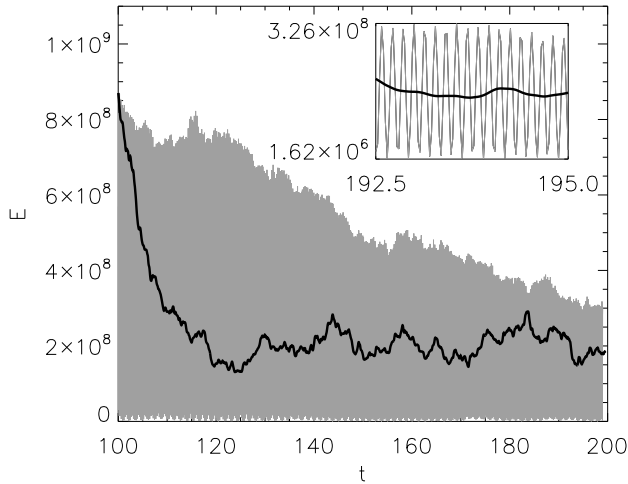}
\caption{Time evolution of the energy in the 2D mode $k=(1,1,0)$ after perturbing the thermalized value by multiplying its amplitude by 10 in runs R14 (thin gray, $\Omega=0$) and R17 (thick black, $\Omega=16$). The inset shows the time evolution for late times; note the oscillations in run R17, corresponding to Rossby waves.}
\label{fig:perturb}
\end{figure}

\subsection{Isotropic statistical equilibrium} \label{ss:spec}

At late times, all the runs (see e.g., Fig.~\ref{fig:en_evol}) reach the statistical equilibrium with zero flux and energy spectrum $\sim k^2$ with no sign of condensation at large scales. To study this later regime in more detail, we focus now on the runs with initial conditions at small scales (runs R14 to R21), as these runs reach thermalization faster. Similar results are obtained at late times in runs R1 to R13.

We show in Fig.~\ref{fig:nohel} the energy spectra for different rotation rates (see caption) at $t= 44 = 220 \tau_0$. Note the evolution of all runs toward a pure $k^2$ spectrum, as expected from the isotropic equations (\ref{eq:equil}). Only the modes at the smallest wavenumbers depart from the equilibrium, as the time it takes for these modes to thermalize increases as $\Omega$ increases. This state is isotropic, as can be confirmed, e.g., by studying the ``reduced'' parallel and perpendicular energy spectra $E(k_\parallel)$ and $E(k_\perp)$ (see e.g., \cite{Alexakis07,Mininni09b}). The spectrum $E(k_\parallel)$ is obtained by integrating the energy over all modes with same wave vector $k_\parallel$ (i.e., integrating over planes in spectral space), and $E(k_\perp)$ is obtained by integrating the energy over all modes with same wave vectors $k_\perp = (k_x^2+k_y^2)^{1/2}$ (i.e, integrating in cylindrical shells). The perpendicular spectrum shows a $\sim k$ dependence (indicated by the solid line for convenience), as expected for equilibria with integration over wavenumbers in one direction only (see also \cite{bourouiba}). The parallel spectrum is flat, again consistent with equipartition of energy between all modes, as in that case all planes in Fourier space contain the same amount of energy. The decrease of both reduced spectra at large wavenumber is due to the fact that, in the anisotropic case, integrations are performed in planes or cylinders but undergo the spherical truncation imposed in the computation (associated to the dealiasing). This results in a deficit of modes at large $|{\bf k}|$ for anisotropic binning of Fourier space.

When helicity is present, the absolute equilibria ensemble are modified as found in \cite{rhk73}, with a behavior close to $k^2$ at low wavenumber but departing from it at high $k$. The result is illustrated in Fig.~\ref{fig:hel}, which shows the reduced energy and helicity spectra in runs R15 and R18 at $t=55=275 \tau_0$. The reduced helicity spectra are constructed following the same procedure as for the energy. Independently of the rotation rate, the spectra are in good agreement with the isotropic theoretical predictions given in \cite{rhk73}. The delay for reaching the asymptotic state is slightly larger in the helical case at the same rotation rate, as can be expected from the fact that for a fully helical Beltrami flow, the Lamb vector ${\bf u} \times \mbox{\boldmath $\omega$} \approx 0$ until the initial flow develops.

In light of these results, it may seem that the final equilibrium is exactly the same for all the runs, independently of the presence of rotation. However, there are differences in the coupling between modes in the rotating case even after thermalization has been achieved. To show this, the final states of runs R14 and R17 were perturbed by multiplying by 10 the amplitudes of the modes $k=(1,1,0)$ (a 2D mode) and $k=(10,10,10)$ (a 3D mode). Return to thermalization of these modes took different times depending on the value of $\Omega$, more clearly in the case of the 2D mode (see Fig.~\ref{fig:perturb}). Moreover, once the thermalized amplitude was reached, the modes in the run without rotation fluctuated in time as white noise, while the same modes in the run with rotation showed clear oscillations with the frequency of the Rossby waves.

\section{Discussion and conclusion\label{s:conclu}}

We have found in the case of helical rotating flows a counter-example to the otherwise well-tested conjecture that statistical ensemble equilibria are good predictors for the small-scale and large-scale behavior of turbulent flows. In three-dimensional helical rotating flows, the equilibrium does not have a condensate with a low wavenumber accumulation of excitation, whereas the dissipative dynamics show a clear inverse cascade. Moreover, the spectra computed for long times for inviscid truncated systems do not show any anisotropization of the large scale. The limitation of the conjecture that statistical ensemble equilibria are good predictors for small- and large-scale behavior was already identified for the case of non-helical rotating flows in \cite{yamazaki,bourouiba}.

One is led to speculate as to what is the origin of this discrepancy; we attribute it to the way the nonlinearity of the system is modified in the presence of inertial waves, as studied from numerical data, e.g., in \cite{Chen03b,Mininni09b}. In rotating turbulence, the development of anisotropy and inverse transfer of energy results from the resonant and near-resonant interactions between three-dimensional modes and modes in the slow manifold with $k_\parallel=0$, per virtue of the linear Coriolis term \cite{Cambon89,Waleffe93}. Energy in three dimensional modes is transferred by a subset of the resonant interactions to modes with smaller vertical wavenumber. It is therefore to be expected that the inverse transfer of energy in a rotating flow will be different from the two-dimensional case (without rotation) insofar as resonances play an essential role. However, the Liouville theorem is not modified by the linear terms, and therefore the statistical equilibrium remains isotropic and without condensates after rotation is introduced, except for early times when the energy in the 2D modes is quasi-conserved. It remains puzzling however to see that the waves, stemming from a linear term in the dynamical equation, perturb a nonlinear effect when those end up dominating the dynamics for sufficiently long times. This paper attempted at elucidating this behavior examining the role that the second invariant, the helicity, has on inviscid dynamics since it is known that it affects the driven dissipative case, making the spectra different \cite{Mininni10a, Mininni10b} from what is expected in the non-helical case \cite{dubrulle, zhou}.

It has been argued \cite{bartello,bourouiba} that, since in the case of very strong rotation, the flow evolves toward a two-dimensional state in which two-dimensional quadratic invariants as the enstrophy are recovered, this can justify the existence of an inverse cascade in the driven dissipative case. Indeed, we can confirm the results of \cite{bourouiba} showing quasi-conservation of two-dimensional invariants for early times. The introduction of helicity further increases the time in which this approximation holds, but keeps as the equilibrium state that of the non-rotating case. However, independently of the presence of helicity, once the isotropic thermalized equilibrium has been reached, differences in the evolution of the modes can be identified. While in the absence of rotation the amplitude of the modes evolve as white noise, traces of rotation and the associated coupling between modes can be identified in the rotating case. Perturbation of any mode requires long times to recover thermalization in the latter case, and time oscillations associated to waves are observed in the amplitude of the modes.

The results are insensitive to the spatial resolution used, at least in the range of values explored here. The time during which $E_\textrm{2D}$ and $E_\textrm{3D}$ are independently conserved seems insensitive to the resolution. Properties of the late time isotropic equilibrium are also unchanged as resolution is increased. However, resolution allows us to resolve better the spectrum that develops during the transient. Our results show, in agreement with previous claims for non-rotating flows \cite{meb}, that during the transition from the approximate conservation of quasi-invariants to the thermalized regime, a dissipative-like spectrum arises. There is a dispute as to what the energy spectrum for rotating flows may be, even when putting aside the effect helicity may have \cite{Mininni10a, Mininni10b}. Weak turbulence theory predicts a $k^{-2}$ spectrum \cite{Galtier03}, consistent with dimensional analysis \cite{dubrulle, zhou} (note that this theory finds at lowest order a decoupling of the 2D and 3D modes). But it is claimed in \cite{cambon2004} that such a spectrum applies only in the vicinity of $k_{\parallel}=0$ and that, as one takes into account the contribution from all parallel wavenumbers, the resulting isotropic spectrum is $\propto k^{-3}$. Note that this power-law should not be viewed as the classical Kraichnan spectrum for two-dimensional Navier-Stokes \cite{rhk_montgo}, but rather as being the isotropic result of an anisotropic distribution of modes taking into account the $k_{\parallel}$ dependence of the dynamics.

Essential to the argument is that the decoupling of 2D and 3D motions on the basis of resonances \cite{Waleffe93} is not complete because of the differential energy between purely 2D and purely 3D vertical velocities in the $k_{\parallel}$ case (the polarization anisotropy $Z_\textrm{PA}$). The results presented here seem to favor this spectrum, with an isotropic scaling $\sim k^{-3}$ which obtains from an anisotropic two-dimensional spectrum $e(k,\cos \theta)$ which scales for $\cos \theta = 1$ as $k^{-2}$ (in the presence of helicity, the isotropic spectrum is even steeper). As a result, the ideal runs seem therefore to be a useful tool to study the high-rotation case in spite of the limitations of the statistical ensemble predictions. It is unclear for the moment why in viscous simulations of non-helical rotating flows the isotropic spectrum scales as $k^{-2}$. It may be the case that dissipation plays an essential role at sufficiently small scale, an effect already noted in \cite{yang_doma} in the decaying case.

\begin{acknowledgments}
The authors are thankful to an anonymous reviewer for a useful remark. Computer time was provided by NCAR and by the University of Buenos Aires. NCAR is sponsored by the National Science Foundation. WHM acknowledges support from NSF grant ATM-0539995; PD and PDM acknowledge support from grants UBACYT X468/08 and X469/08, PICT 2007-02211 and 2007-00856.
\end{acknowledgments}

\end{document}